# A double quantum dot defined by top gates in a single crystalline InSb nanosheet

Yuanjie Chen[1], Shaoyun Huang[1], Jingwei Mu[1], Dong Pan[2,3], Jianhua Zhao[2,3], and H. Q. Xu*[1,3]

[1] Beijing Key Laboratory of Quantum Devices, Key Laboratory for the Physics and Chemistry of Nanodevices and Department of Electronics, Peking University, 100871 Beijing, China.

[2] State Key Laboratory of Superlattices and Microstructures, Institute of Semiconductors, Chinese Academy of Sciences, P.O. Box 912, 100083 Beijing, China.

[3] Beijing Academy of Quantum Information Sciences, 100193 Beijing, China.

*Email: hqxu@pku.edu.cn

(Dated: May 8th, 2021)

**Keywords**: two-dimensional materials, InSb nanosheet, quantum dot

**PACS**: 85.35.Be; 73.63.Kv; 73.63.-b

## Abstract

We report on the transport study of a double quantum dot (DQD) device made from a freestanding, single crystalline InSb nanosheet. The freestanding nanosheet is grown by molecular beam epitaxy and the DQD is defined by top gate technique. Through the transport measurements, we demonstrate how a single quantum dot (QD) and a DQD can be defined in an InSb nanosheet by tuning voltages applied to the top gates. We also measure the charge stability diagrams of the DQD and show that the charge states and the inter-dot coupling between the two individual QDs in the DQD can be efficiently regulated by the top gates. Numerical simulations for the potential profile and charge density distribution in the DQD have been performed and the results support the experimental findings and provide a better understanding of fabrication and transport characteristics of the DQD in the InSb nanosheet. The achieved DQD in the two-dimensional InSb nanosheet possesses pronounced benefits in lateral scaling and can thus serve as a new building block for developments of quantum computation and quantum simulation technologies.

## 1. Introduction

Narrow bandgap InSb nanostructures with strong spin-orbit coupling (SOC) and large Landé g-factor have attracted growing attention in the fields of spintronics and quantum information science and technology. Spin-based quantum states—especially in InSb quantum dots (QDs)—can be stored and manipulated by fast, efficient, and all-electrical means[1,2]. Two-dimensional (2D) multiple QD structures not only can be employed as building blocks for constructing semiconductor QD qubits[3,4] and quantum simulators[5,6], but also can help to read out[7,8] and braid[9,10] Majorana zero modes in semiconductor-based topological superconducting devices. However, advances in the realization and applications of QDs made from 2D InSb systems have been impeded due to the lack of an established fabrication technique. Charge instabilities, gate hysteresis and leakage problems [11-13] have limited the realization and investigations of high-quality, confined systems in heterostructured InSb quantum wells capped with thick buffer layers [14-16]. Recently, by removing the doping layer from an InSb heterostructure and by inducing carriers to the InSb quantum well electrostatically, a progress in the realization of stable gate-defined QDs in an InSb quantum well has been achieved [17]. An alternative approach is to employ a free-standing InSb nanolayer, i.e., a single crystalline InSb structure without any other material layers and doping layers around. Such free-standing, zincblende InSb nanosheets [18] have recently been grown via molecular beam epitaxy (MBE). Electrical measurements have shown that these freestanding nanosheets exhibit good transport properties and tunable, strong SOC, and are able to be contacted directly via metal deposition. In a recent report, experimental attempts to define a single QD in an MBE-grown freestanding InSb nanosheet have been made via top gate technique [19]. Although signs of quantum confinement are observed in the transport measurements of the device, it is hard to achieve desired high-quality charge state stability diagrams and good controllability in the charge states of the QD defined in the InSb nanosheet, thus casting a serious question about whether these freestanding nanosheets could be employed to develop integrated QD devices for quantum information science and technology applications.

In this letter, we report on the realization of a good-quality double QD (DQD) in an MBE-grown freestanding InSb nanosheet. The DQD confinement is achieved all by top finger gates, with a thin (10 nm) layer of $HfO_2$ as gate dielectric, using a gate layout optimized for InSb nanosheets. The fabricated devices are measured in a

$^3$He/$^4$He dilution refrigerator with a base temperature of $T \sim 20$ mK. Regular Coulomb diamonds are observed in the charge stability diagram of each individual QD, showing that the QDs can now be much better defined in the InSb nanosheet than that in Ref. 19. The measured charge stability diagrams of the DQD demonstrate that the charge states and the inter-dot coupling of the DQD can be efficiently regulated by the top gates. We also perform a numerical simulation based on COMSOL program for the DQD. The results support our measurements and provide a better understanding of the fabrication and transport characteristics of the DQD made in the InSb nanosheet.

## 2. Experiment
### 2.1 Materials

The planar DQD device reported in this work is made from a free-standing, single crystalline InSb nanosheet. Freestanding InSb nanosheets are grown via MBE on top of individual InAs nanowire stems on a Si(111) substrate. Prior to the growth process, a thin film of Ag is deposited onto the Si substrate in the MBE chamber and is annealed *in situ* to form Ag nanoparticles. With the use of these Ag nanoparticles as seeds, InAs nanowires are first grown on the Si substrate. By abruptly switching the group-V source from As to Sb and gradually increasing the supplied flux of Sb, freestanding InSb nanosheets are formed on top of these InAs nanowires. The grown InSb nanosheets are investigated by scanning electron microscopy and transmission electron microscopy measurements. It is found that the morphology and size of the InSb nanosheets can be controlled by tailoring the Sb/In beam pressure ratio and growth time. It is also seen that these nanosheets are pure-phase zincblende crystals with flat surfaces, free from stacking faults, and of rhombic shapes with lateral sizes up to several micrometers and a thickness ranging from 10 to 80 nm. For further details about the growth and structural properties as well as basic transport properties of our MBE-grown InSb nanosheets, we would refer to Ref. 18, 20, and 21.

### 2.2 Device fabrication and characterization

For device fabrication, the MBE-grown InSb nanosheets are mechanically transferred onto a heavily doped Si substrate (employed as a global back gate to fabricated nanosheet devices) with a 300-nm-thick capping layer of SiO$_2$, and predefined metal markers on top. Subsequent to the nanosheet transfer, electron beam lithography (EBL) is carried out to define the source and drain contact regions on selected InSb nanosheets. The metal contacts are then fabricated by means of electron beam evaporation (EBE) of 5-nm-thick Ti and 90 nm-thick Au, and a lift-off process. In order to obtain good contacts to the InSb nanosheets, the samples are wet etched in a de-ionized water-diluted (NH$_4$)$_2$S$_x$ solution at a temperature of 40 °C for several minutes for removal of the native oxide layer and for surface passivation in the opening regions of the InSb nanosheets before being loaded into the EBE chamber. Then the nanosheets are capsulated by a 10-nm-thick layer of HfO$_2$ (a top gate dielectric layer) through a combination of EBL, atomic layer deposition, and lift-off process. Finally, the patterned top gates are fabricated on top of the HfO$_2$ layer by a third step of EBL, followed by EBE of 5-nm-thick Ti and 90-nm-thick Au, and lift-off. Figure 1(a) shows the false-colored scanning electron microscopy image of a typical DQD device made from an InSb nanosheet with thickness $t \sim 30$ nm. In the image, the scale bar is 500 nm, the source and drain, denoted by S and D, are colored in yellow, while the top gates, denoted by G1 to G6, are colored in light purple. Here, top gates G1 to G6 all are employed to provide electrostatic confinement to electrons in the InSb nanosheet. But, top gates G3 and G5 are also employed as plunger gates to control the chemical potential of the two individual QDs in the DQD, and top gate G4 is employed to tune the inter-dot coupling between the two individual QDs. Now, it is important to note that in a traditional 2D electron gas system made from a heterostructure with a thick cap layer, top gate electrodes are designed with a relatively large spacing between them to produce an appropriate confinement potential profile[22,23]. However, this might not be suitable for InSb materials with a very narrow bandgap (~0.23 eV). To realize complete depletion of electrons and at the same time prevent hole accumulation[24] in an InSb nanosheet, we have optimized the layout of top gate electrodes and the dielectric layer. On one hand, the dielectric layer should be thin enough such that a smaller negative gate voltage can deplete the local area of conducting electrons. On the other hand, the spacing between top gate electrodes should be narrow enough to provide sufficient confinement for electrons without the need of applying a large negative voltage. In this work, the spacing between gates G1 and G2 is designed to be 50 nm, and the same as the one between gates G1 and G6. The small top gate spacing combined with a thin dielectric layer is essential for an accurate definition of electron confinement without inducing hole accumulation in an InSb nanosheet.

The fabricated devices are measured in a $^3$He/$^4$He dilution refrigerator at a base temperature of $T \sim 20$ mK. In order to improve the signal-to-noise ratio, electronic noises in the measurement circuits are minimized using a homemade copper-powder filter and a multiple RC filter set at different temperature stages. Below, we report the results of our measurements for the DQD device shown in Fig. 1(a).

## 3. Results and discussion
### 3.1 Back-gate characterization

We first evaluate the back-gate performance of the DQD device depicted in Fig. 1(a). Figure 1(b) shows the measured source-drain current $I_{ds}$ as a function of back-gate voltage $V_{bg}$ (gate transfer characteristics) at temperature $T \sim 20$ mK and source-drain bias voltage $V_{ds} = 0.5$ mV with all the top gates being grounded (i.e., by setting all the back-gate voltages at $V_{G1\text{-}G6} = 0$ V). Evidently, this device shows a typical n-type carrier transport characteristic with a pinch-off threshold back-gate voltage of $V_{bg}^{th} \sim -3.5$ V. The observed current fluctuations superimposed on the gate transfer curve are reproducible and can be attributed to interference effects such as universal conductance fluctuations. Taking the influence of finite contact resistance $R_c$ into account, the source-drain current $I_{ds}$ can be expressed as

$$I_{ds} = \frac{V_{ds}}{G_s^{-1} + 2R_c}. \tag{1}$$

Here, $G_s$ is the channel conductance and can be expressed as $G_s = \frac{\mu}{L^2} C_{bg}(V_{bg} - V_{bg}^{th})$, where $\mu$ denotes the field-effect electron mobility, $L \sim 1.1$ $\mu$m is the length of the conduction channel, i.e., the spacing between source and drain electrodes, and $C_{bg}$ is the capacitance of the back gate to the InSb nanosheet channel. Based on the parallel plate capacitor model, $C_{bg}$ can be estimated from $C_{bg} = \frac{\varepsilon_0 \varepsilon_r S}{d} \sim 1.1 \times 10^{-16}$ F, with $\varepsilon_0$ denoting the vacuum permittivity, $\varepsilon_r \sim 3.9$ the relative permittivity of SiO$_2$, $d \sim$ 300 nm the thickness of SiO$_2$ layer, and $S \sim 0.93$ $\mu$m$^2$ the area of the InSb nanosheet channel. The mobility $\mu$, threshold $V_{bg}^{th}$, and contact resistance $R_c$ can be obtained by fitting the measured transfer curve to Eq. (1). The fitting yields $\mu = 1.41 \pm 0.03$ m$^2$V$^{-1}$s$^{-1}$, $V_{bg}^{th} = -3.46 \pm 0.01$ V, and $R_c = 1.78 \pm 0.02$ k$\Omega$ at $V_{ds} = 0.5$ mV. The extracted mobility exceeds $\sim 1$ m$^2$V$^{-1}$s$^{-1}$, which together with the small contact resistance [25] shows good transport properties of the InSb nanosheet. In the following measurements, we set the back-gate voltage at $V_{bg} = 1$ V [as marked by a yellow star in Fig. 1(b)]. The corresponding sheet carrier density can be estimated as $n = \frac{C_{bg}}{S} \times \frac{V_{bg} - V_{bg}^{th}}{e} = 3.21 \times 10^{11}$ cm$^{-2}$, with e being the elementary charge. The mean free path can be determined as $l_e = \frac{\hbar \mu}{e} \sqrt{2\pi n} \sim 130$ nm, comparable to the size of $\sim 150$ nm designed for individual QDs in the device, where $\hbar$ is the reduced Planck constant. Thus, at this sufficiently high carrier density, influences of disorder-induced potential fluctuations are negligibly weak in the definition of the DQD in the InSb nanosheet [26,27].

**3.2 Top-gate characterizations**

In this subsection, we determine the performance and setting point of top finger gates in the device. Figure 2(a) shows the measured $I_{ds}$ as a function of $V_{G1}$, i.e., the voltage applied to gate G1, at $V_{ds} = 0.5$ mV and $V_{bg} = 1$ V, with all other top gates being grounded (i.e., $V_{G2-G6} = 0$ V). As $V_{G1}$ decreases from 0 V to about $-0.5$ V, $I_{ds}$ gets suppressed due to the depletion of the conducting carriers right beneath gate G1. When $V_{G1}$ continues to decrease, the current $I_{ds}$ settles at an approximately constant value of $\sim 40$ nA, indicating that the negative voltage $V_{G1}$ has totally depleted the carriers in the region covered by gate G1 but nearly has no influence on the region which are not covered by gate G1. We then set $V_{G1} = -1.2$ V [as marked by the yellow star in Fig. 2(a)] and check the operating characteristics of other top gates. Figure 2(b) shows the measured current $I_{ds}$ as a function of $V_{G2}$, i.e., the voltage applied to gate G2, at $V_{ds} = 0.5$ mV, $V_{bg} = 1$ V, and $V_{G1} = -1.2$ V, with all the remaining top gates being grounded. It is seen that the current $I_{ds}$ gets pinched off at $V_{G2} = -0.75$ V, showing that a quantum point contact (QPC) structure that consists of gates G1 and G2 functions efficiently to form a potential barrier to the conducting electrons in the nanosheet. In a similar way, the operating characteristics of top gates G4 and G6 are evaluated. As shown in Figs. 2(c) [2(d)], the measured current $I_{ds}$ as a function of the voltage applied to gate G4 (G6) approaches to zero at $V_{G4} = -0.53$ V ($V_{G6} = -0.65$ V), indicating that the formations of two QPCs, with one by gates G1 and G4 and the other one by gates G1 and G6, are achieved in the nanosheet. Note that the pinch-off voltages of the top gates G2, G4, and G6 $> -1$ V, which are much smaller than that in Ref. 19. This can be attributed to the thinner $HfO_2$ layer and optimized layout of the top gate electrodes. Based on these measured operating characteristics, we have set in defining our DQD the voltages applied the three finger top gates at $V_{G2} = -0.74$ V, $V_{G4} = -0.5$ V and $V_{G6} = -0.63$ V, which are slightly less negative than their own pinch-off thresholds in order to compensate the cross-talk effects of remote gates on these QPCs. The remaining top gates G3 and G5 are also characterized and are found to function properly. These two top gated are employed as plunger gates to the DQD. Note that the resonance-like fluctuation structures seen in all the measured current curves shown in Fig. 2 are reproducible. Such fluctuations are commonly observed in the measurements of quantum devices made from free-standing semiconductor nanostructures, such as InAs and InSb nanowires. These fluctuations are inherent to a mesoscopic system in which the electron coherence length is on the order of the size of the conduction channel and the electron mean free path is much shorter than the conduction channel size[19].

## 3.3 Defining and transport characteristics of the DQD

In this subsection, we report our transport measurements of the DQD defined in the InSb nanosheet as described above. We first demonstrate the formation of two individual QDs which constitute the DQD. Figure 3(a) shows the measured current $I_{ds}$ through the left QD, defined as in Fig. 1(a) by setting top-gate voltages at $V_{G1} = -1.2$ V, $V_{G2} = -0.74$ V, and $V_{G4} = -0.5$ V, as a function of the plunger gate voltage $V_{G3}$ at source-drain bias voltage $V_{ds} = 0.1$ mV. Here, in the measurements, top gates G5 and G6 are grounded, i.e., $V_{G5}$ and $V_{G6}$ are set at $V_{G5} = V_{G6} = 0$ V. Voltage applied to the back gate is always set at $V_{bg} = 1$ V through the entire measurements reported in this subsection. In Fig. 3(a), consecutive sharp current peaks (Coulomb current oscillation peaks) are observed over a wide range of $V_{G3}$, providing clear evidence of single electron transport through a QD in the Coulomb blockade regime[28]. Figure 3(c) shows the measured differential conductance $dI_{ds}/dV_{ds}$ of the left QD as a function of bias voltage $V_{ds}$ and gate voltage $V_{G3}$ (charge stability diagram). Here, we see a series of regular Coulomb diamond structures with closed degeneracy points at zero bias between neighboring Coulomb diamonds, indicating the formation of a well-defined single QD. It is also observed that the size of Coulomb diamonds (or the single-electron addition energy of the QD) shows a deviation from the regular even-odd effect which has often been observed in the few electron regimes of a QD due to the spin degeneracy. This deviation is most likely due to the presence of double level degeneracy in our laterally defined QD[29-32]. In the following, this left QD formed by G1, G2, G3, and G4 is denoted as QD1. Figure 3(b) shows the measured current $I_{ds}$ through the right QD (to be denoted as QD2), defined as in Fig. 1(a) by setting top-gate voltages at $V_{G1} = -1.2$ V, $V_{G4} = -0.5$ V, and $V_{G6} = -0.63$ V, as a function of the voltage $V_{G5}$ applied to plunger gate G5 at source-drain bias voltage $V_{ds} = 0.1$ mV. Here, the two uninvolved top gates G2 and G3 are grounded, i.e., $V_{G2} = 0$ V and $V_{G3} = 0$ V are set in the measurements. The inset in Fig. 3(b) displays the zoom-in plot of two tiny current peaks in the gate voltage region marked by the red rectangle in the main figure. Clearly, the measured source-drain current $I_{ds}$ exhibits Coulomb oscillations, i.e., signatures of single electron transport through a QD. Figure 3(d) shows the differential conductance $dI_{ds}/dV_{ds}$ measured for QD2 as a function of $V_{ds}$ and $V_{G5}$ (charge stability diagram). Here, successive regular Coulomb diamonds are again observed, indicating that QD2 is well defined in the InSb nanosheet. Note that an extra, weak differential conductance line, marked by a yellow arrow in Fig. 3(d), with a different slope from that of the boundaries of the Coulomb diamonds is observable. This line could most likely arise from a localized or impurity state formed unintentionally in a remote place from gate G5 and thus has a weaker coupling to the gate[33]. Note that noticeable differences can be identified in the measured Coulomb current oscillations and charge stability diagrams of QD1 and QD2. These differences can be attributed to the differences in the sizes of confinement of the two QDs and indicate that deviations from our designed layout are present in our fabricated gate structures.

Having confirmed the two individually controllable single QDs, we now demonstrate the formation and transport properties of the DQD defined as in Fig. 1(a) by setting $V_{G1} = -1.2$ V, $V_{G2} = -0.74$ V, $V_{G4} = -0.473$ V, $V_{G6} = -0.63$ V, and $V_{bg} = 1$ V. Figure 4(a) shows the measured source-drain current $I_{ds}$ through the DQD at a source-drain bias voltage of $V_{ds} = 0.1$ mV as a function of plunger gate voltages $V_{G3}$ and $V_{G5}$ (charge stability diagram of the DQD). It is seen that the charge stability diagram is featured by a chessboard-like pattern[34], i.e., a group of relatively strong vertical current lines and a group of rather weak horizontal current lines (marked by horizontal dashed lines), and by small or hardly visible interactions between the two groups of current lines. The observation of these vertical and horizontal current lines indicates that plunger gates G3 and G5 address only effectively their own targeting QDs and have very small crosstalk capacitances to the other QDs. The weak interactions between the vertical and horizontal current lines indicate that the DQD is in the weak inter-dot coupling regime. Figure 4(b) displays the measured charge stability diagram of the DQD defined with the gate voltages all set the same as in Fig. 4(a) except for the voltage applied to the inter-dot coupling gate G4 which has been set at $V_{G4} = -0.42$ V. As $V_{G4}$ increases from $-0.473$ V to $-0.42$ V, the coupling between the two QDs in the DQD and the crosstalk capacitances between the plunger gates and the QDs other than their locally addressing ones are enhanced. These enhancements lead to appearances of finite slopes in the two groups of the current lines and of anti-crossing characteristics at the crosses of the two groups of the current lines. As a result, the current lines tend to show a honeycomb pattern in the charge stability diagram, see, e.g., a hexagon highlighted by the white dashed rectangle. In the lowest order transport process, the current flow occurs only at the corners, i.e., triple points, of the hexagons[35], while the higher-order co-tunneling processes give rise to the current observed at the boundaries of the hexagons[34]. Figure 4(c) shows the measured charge stability diagram of the DQD at $V_{G4} = -0.35$ V, where the DQD is in the strong inter-dot coupling regime. As shown in Fig. 4(c), the charge stability diagram is characterized by a series of diagonal current lines (with negative slopes), reminiscent of the formation of a large single QD. The bending of the straight current lines occurs when quantum levels in the two dots move close in energy. So far, we have demonstrated the realization of a DQD in our InSb nanosheet and that it is possible to tune the DQD from a weak inter-dot coupling regime to an intermediate and finally a strong inter-dot coupling regime with tuning plunger gate voltage $V_{G4}$.

## 3.4 Simulation of the charge density and confinement potential in the DQD device

To get a better understanding of the formation of the DQD in the InSb nanosheet, we perform simulation for the charge density and electrostatic potential distributions in the system via finite element method using commercially available program COMSOL[36,37]. Figures 5(a)-5(c) show the simulated 2D charge density profile $n^{2D}_{simu}$ in the InSb layer of the DQD system for the three experimental cases as in the Figs. 4(a)-4(c). In the simulation, the same layer structure and device layout as in the experiment are employed. The average sheet carrier density $n^{2D} = 3.21 \times 10^{11}$ cm$^{-2}$ extracted from the measured back-gate transfer characteristics (i.e., the $I_{ds}$-$V_{bg}$ curve) is employed. The Fermi energy of the electrons in the conducting channel, free of the impact of the top gates, is then estimated as $E_F = \frac{n\pi\hbar^2}{m^*} \approx 51$ meV, where $m^* = 0.014\ m_0$ denotes the electron effective mass of InSb and $m_0$ is the mass of a free electron. The simulation starts with the determination of the spatial distribution of electric potential $\phi$ inside the InSb layer using the three-dimensional (3D) Poisson's equation,

$$\nabla^2 \phi = \frac{e}{\epsilon} n^{3D}, \qquad (2)$$

with the same electrostatic boundary conditions, e.g., the gate voltages, as in the experiment. Here, $\epsilon$ is the dielectric constant and the initial volume carrier density $n^{3D} = 1.07 \times 10^{17}$ cm$^{-3}$ without considering the impact of the top gates is estimated out from the measured sheet carrier density $n^{2D}$ by assuming that the carriers are uniformly distributed in the InSb nanosheet. Then, to the lowest order of Thomas-Fermi approximation[38,39], the local sheet charge density $n^{2D}_{simu}$ under different sets of the voltages applied to the top-gates is calculated as

$$n^{2D}_{simu} = \int_Z dz (E_F + e\phi) / \frac{\pi\hbar^2 t}{m^*}, \qquad (3)$$

where $t = 30$ nm is the thickness of the InSb nanosheet and the integration is performed along the direction perpendicular to the InSb plane. Here, we define the direction perpendicular to the InSb plane as the $z$-direction, and the $x$-$y$ plane is defined to be parallel to the InSb plane with the $x$-axis ($y$-axis) being along (perpendicular to) the direction of current flow. It is seen in Figs. 5(a)-5(c) that the simulation demonstrates good confinement of electrons by the top gates and thus the formation of the DQD in the InSb nanosheet. It is also seen that as $V_{G4}$ increases, the maximum charge density inside the DQD increases gradually from ~2×10$^{11}$ cm$^{-2}$ to ~2.4 ×10$^{11}$ cm$^{-2}$ and the charge population in the joint region between the two composite QDs increases from nearly zero to ~1.5×10$^{11}$ cm$^{-2}$. The results indicate that with increasing $V_{G4}$, the couplings between the higher-energy electron states in the two QDs would become stronger.

Figures 5(d)-5(f) show the calculated electron potential energy $-e\phi$ along the line cut of $y = 0$, $z = 0$, i.e., across the centers of the two QDs inside the InSb nanosheet, for the cases corresponding to Figs. 5(a)-5(c). The red dashed lines in the figures mark the Fermi energy $E_F \sim 0.051$ eV in the InSb layer. It is seen that the potential barrier height between the two QDs varies from ~0.070 eV above the Fermi energy to ~0.039 eV below the Fermi energy, giving an increase in coupling between QD1 and QD2 and thus supporting our experimental observation that the DQD has undergone a weak-to-strong inter-dot coupling transition with increasing $V_{G4}$.

**4. Conclusions:**

In this work, a DQD device defined by optimized top gates is realized in an emerging high-quality, narrow bandgap, semiconductor InSb nanosheet. The fabricated device is studied by transport measurements in a $^3$He/$^4$He dilution refrigerator at a base temperature of $T \sim 20$ mK. The charge stability diagrams are measured for the DQD in different inter-dot coupling regimes. It is shown that both the charge states and the inter-dot coupling of the DQD can be efficiently tuned using the designed, dedicated top gates. Numerical simulation of the electrical potential and charge density distributions in the DQD have been performed and the results support the experimental observations and provide an insight into the structure design and performance of the DQD device. Our work sets a solid experimental step towards the development of laterally integrated semiconductor QD structures for quantum simulation and quantum computing.

**Acknowledgment**
This work is supported by the Ministry of Science and Technology of China through the National Key Research and Development Program of China (Grant Nos. 2017YFA0303304, 2016YFA0300601, 2017YFA0204901, and 2016YFA0300802), the National Natural Science Foundation of China (Grant Nos. 91221202, 91421303, 11874071, 11974030 and 61974138), the Beijing Academy of Quantum Information Sciences (No. Y18G22), the Key-Area Research and Development Program of Guangdong Province (Grant No. 2020B0303060001), and the Beijing Natural Science Foundation (Grant Nos. 1202010 and 1192017). DP also acknowledges the support from Youth Innovation Promotion Association, Chinese Academy of Sciences (No. 2017156).

**Reference**

1.  Nadj-Perge S, Frolov S M, Bakkers E P A M and Kouwenhoven L P 2010 *Nature* **468** 1084

2.  Nadj-Perge S, Pribiag V S, Van den Berg, J W G, Zuo K, Plissard S R, Bakkers E P A M, Frolov S M and Kouwenhoven L P 2012 *Phys. Rev. Lett.* **108** 166801

3.  Loss D and DiVincenzo D P 1998 *Phys. Rev. A* **57** 120

4.  Zhang X, Li H-O, Cao G, Xiao M, Guo G-C and Guo G-P 2019 *Natl. Sci. Rev.* **6** 32

5.  Georgescu L M, Ashhab S and Nori F 2014 *Rev. Mod. Phys.* **86** 153

6.  Hensgens T, Fujita T, Janssen L, Li X, Van Diepen C J, Reichl C, Wegscheider W, Sarma S D and Vandersypen L M K 2017 *Nature* **548** 70

7.  Aasen D, Hell M, Mishmash R V, Higginbotham A, Danon J, Leijnse M, Jespersen T S, Folk J A, Marcus C M, Flensberg K and Alicea J 2016 *Phys. Rev. X* **6** 031016

8.  Gharavi K, Hoving D and Baugh J 2016 *Phys. Rev. B* **94** 155417

9.  Malciu C, Mazza L and Mora C 2018 *Phys. Rev. B* **98** 165426

10. Zhou Y-F, Hou Z and Sun Q-F 2019 *Phys. Rev. B* **99** 195137

11. Yi W, Kiselev A A, Thorp J, Noah R, Nguyen B-M, Bui S, Rajavel R D, Hussain T, Gyure M F, Kratz P, Qian Q, Manfra M J, Pribiag V S, Kouwenhoven L P, Marcus C M and Sokolich M 2015 *Appl. Phys. Lett.* **106** 142103

12. Uddin M M, Liu H W, Yang K F, Nagase K, Sekine K, Gaspe C K, Mishima T D, Santos M B and Hirayama Y 2013 *Appl. Phys. Lett.* **103** 123502

13. Orr J M S, Buckle P D, Fearn M, Wilding P J, Bartlett C J, Emeny M T, Buckle L

and Ashley T 2006 *Semicond. Sci. Technol.* **21** 1408

14. Orr J M S, Buckle P D, Fearn M, Storey C J, Buckle L and Ashley T 2009 *New J. Phys.* **9** 261

15. Qu F, Veen J V, De Vries F K, Beukman A J A, Wimmer M, Yi W, Kiselev A A, Nguyen B-M, Sokolich M, Manfra M J, Nichele F, Marcus C M and Kouwenhoven L P 2016 *Nano Lett.* **16** 7509

16. Masuda T, Sekine K, Nagase K, Wickramasinghe K S, Mishima T D, Santos M B and Hirayama Y 2018 *Appl. Phys. Lett.* **112** 192103

17. Kulesh I, Ke C T, Thomas C, Kaewal S, Moehle C M, Metti S, Kallaher R, Gardner G C, Manfra M J and Goswami S 2020 *Phys. Rev. Appl.* **13** 041003

18. Pan D, Fan D X, Kang N, Zhi J H, Yu X Z, Xu H Q and Zhao J H 2016 *Nano Lett.* **16** 834

19. Xue J, Chen Y, Pan D, Wang J-Y, Zhao J, Huang S and Xu H Q 2019 *Appl. Phys. Lett.* **114** 023108

20. Kang N, Fan D, Zhi J, Pan D, Li S, Wang C, Guo J, Zhao J H and Xu H Q 2019 *Nano Lett.* **19** 561

21. Chen Y, Huang S, Pan D, Xue J, Zhang L, Zhao J and Xu H Q 2021 *npj 2D Mater. Appl.* **5** 3

22. Petta J R, Johnson A C, Taylor J M, Laird E A, Yacoby A, Lukin M D, Marcus C M, Hanson M P and Gossard A C 2005 *Science* **309** 2180

23. Shulman M D, Dial O E, Harvey S P, Bluhm H, Umansky V and Yacoby A 2012 *Science* **336** 202

24. Uddin M M, Liu H W, Yang K F, Nagase K, Mishima T D, Santos M B and Hirayama Y 2012 *Appl. Phys. Lett.* **101** 233503

25. Fan D, Kang N, Ghalamestani S G, Dick K A and Xu H Q 2016 *Nanotechnology* **27** 275204

26. Pisoni R, Lei Z, Back P, Eich M, Overweg H, Lee Y, Watanabe K, Taniguchi T, Ihn T and Ensslin K 2018 *Appl. Phys. Lett.* **112** 123101

27. Wang K, De Greve K, Jauregui L A, Sushko A, High A, Zhou Y, Scuri G,

Taniguchi T, Watanabe K and Lukin M D 2018 *Nat. Nanotechnol.* **13** 128

28. Sun J, Larsson M, Maximov I, Hardtdegen H and Xu H Q 2009 *Appl. Phys. Lett.* **94** 042114

29. Tarucha S, Austing D G, Honda T, van der Hage R J and Kouwenhoven L P 1996 *Phys. Rev. Lett.* **77** 3613

30. Nagaraja S, Matahne P, Thean V-Y, Leburton J-P, Kim Y-H and Martin R M 1997 *Phys. Rev. B* **56** 15752

31. Larsson M, Hardtdegen H, Nilsson H A and Xu H Q 2009 *Appl. Phys. Lett.* **95** 192112

32. Hofmann A, Maisi V F, Gold C, Krähenmann T, Rössler C, Basset J, Märki P, Reichl C, Wegscheider W, Ensslin K and Inn T 2016 *Phys. Rev. Lett.* **117** 206803

33. Song X-X, Liu D, Mosallanejad V, You J, Han T-Y, Chen D-T, Li H-O, Cao G, Xiao M, Guo G-C and Guo G-P 2015 *Nanoscale* **7** 16867

34. Van der Wiel W G, De Franceschi S, Elzerman J M, Fujisawa T, Tarucha S and Kouwenhoven L P 2002 *Rev. Mod. Phys.* **75** 1

35. Hanson R, Kouwenhoven L P, Petta J R, Tarucha S and Vandersypen M K 2007 *Rev. Mod. Phys.* **79** 1217

36. Zajac D M, Hazard T M, Mi X, Nielsen E and Petta J R 2016 *Phys. Rev. Appl.* **6** 054013

37. Hamer M, Tóvári E, Zhu M, Thompson M D, Mayorov A, Prance J, Lee Y, Haley R P, Kudrynskyi Z R, Patanè A, Terry D, Kovalyuk Z D, Ensslin K, Kretinin A V, Geim A and Gorbachev R 2018 *Nano Lett.* **18** 3950

38. Tang T-W, O'Regan T and Wu B 2004 *J. Appl. Phys.* **95** 7990

39. Pino R 1998 *Phys. Rev. B* **58** 4644

**Captions:**

**Fig. 1 Device structure and back-gate transfer characteristics** (a) False-colored scanning electron microscope image of a typical DQD device studied by transport measurements in this work. The scale bar is 500 nm. (b) Measured source-drain current $I_{ds}$ at $V_{ds}$ = 0.5 mV as a function of back-gate voltage $V_{bg}$ (the blue solid line) and the results of fit of the measurements to Eq. (1) (the red solid line). The yellow star marks the setting point of the back-gate voltage at $V_{bg}$ = 1 V for the rest measurements reported in this work.

**Fig. 2 Transfer characteristics of the top gates.** (a) Measured $I_{ds}$ at $V_{ds}$ = 0.5 mV, $V_{bg}$ = 1 V, as a function of $V_{G1}$ with all other top gates grounded, i.e., $V_{G2\text{-}G6}$ = 0 V. (b) Measured $I_{ds}$ at $V_{ds}$ = 0.5 mV, $V_{bg}$ = 1 V, $V_{G1}$ = −1.2 V, as a function of $V_{G2}$ with remaining top gates grounded, i.e., $V_{G3\text{-}G6}$ = 0 V. (c) The same as (b) but for $I_{ds}$ as a function of $V_{G4}$ at $V_{G1}$ = −1.2 V, $V_{G2}$ = $V_{G3}$ = $V_{G5}$ = $V_{G6}$ = 0 V. (d) The same as (b) but for $I_{ds}$ as a function of $V_{G6}$ at $V_{G1}$ = −1.2 V, $V_{G2\text{-}G5}$ = 0 V. The yellow stars in the four panels mark the setting points of the voltages applied to the these top gates, i.e., $V_{G1}$ = −1.2 V, $V_{G2}$ = −0.74 V, $V_{G4}$ = −0.5 V, and $V_{G6}$ = −0.63 V.

**Fig. 3 Formation and characterization of two individual QDs.** (a) Measured $I_{ds}$ as a function of $V_{G3}$ at $V_{ds}$ = 0.1 mV for the left QD (QD1) in Fig. 1(a) defined by setting $V_{G1}$ = −1.2 V, $V_{G2}$ = −0.74 V, and $V_{G4}$ = −0.5 V. The current oscillation peaks represent the single electron tunneling processes through QD1. (b) Measured $I_{ds}$ as a function of $V_{G5}$ at $V_{ds}$=0.1 mV for the right QD (QD2) defined by setting $V_{G1}$ = −1.2 V, $V_{G4}$ = −0.5 V, and $V_{G6}$ = −0.63 V. The inset is the zoom-in plot of the region marked by a red rectangle in the figure. The current oscillation peaks represent the single electron tunneling processes through QD2. (c) Measured differential conductance $dI_{ds}/dV_{ds}$ (in a logarithmic scale) as a function of bias voltage $V_{ds}$ and plunger gate voltage $V_{G3}$ for QD1 (charge stability diagram of QD1). The setup of the top gate voltages is the same as in (a). (d) Measured charge stability diagram of QD2. The setup of the top gate voltages is the same as in (b). The yellow arrow marks the appearance of an unintended differential conductance line as discussed in the text.

**Fig. 4 Charge stability diagrams of the DQD in three different inter-dot coupling regimes.** (a)-(c) Measured $I_{ds}$ as a function of voltages $V_{G3}$ and $V_{G5}$ applied to plunger gates G3 and G5 (charge stability diagram) of the DQD, defined by setting $V_{G1} = -1.2$ V, $V_{G2} = -0.74$ V, and $V_{G6} = -0.63$ V, in the three different inter-dot coupling regimes defined by setting $V_{G4}$ at $-0.473$ V, $-0.42$ V, and $-0.35$ V, respectively. The yellow dashed lines in (a) mark the weak, nearly invisible horizontal current lines.

**Fig. 5 Simulated charge density distribution and electron confinement potential in the DQD structure.** (a)-(c) Simulated 2D electron density profile $n_{simu}^{2D}$ in the DQD for the three inter-dot coupling regimes as in Figs 3(a)-(3c). Here, we present a 40°-tilted view of the plots with the color-coded 2D electron density $n_{simu}^{2D}$ represented also by the height in the z direction. (d)-(f) Simulated electron potential energy $-e\phi$ as a function of the $x$ coordinate at $y = 0$ and $z = 0$, i.e., a line-cut across the centers of the two QDs in the InSb nanosheet. The red dashed lines mark the Fermi energy $E_F \sim 0.051$ eV in the layer free from the impact of the top gates.

**Figures:**

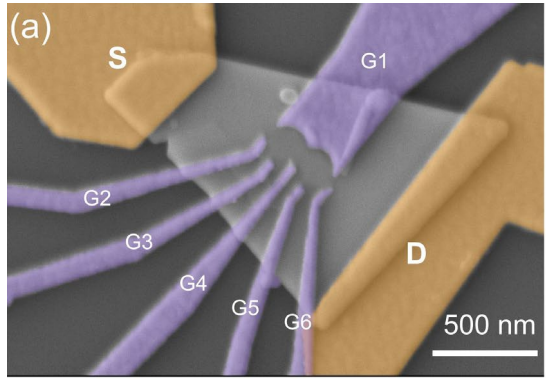 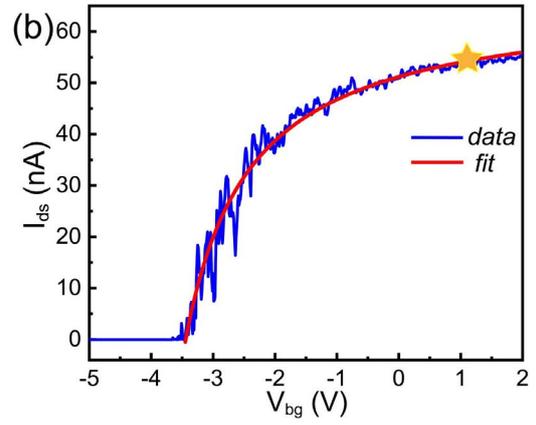

**Fig. 1 by Chen Y *et al*.**

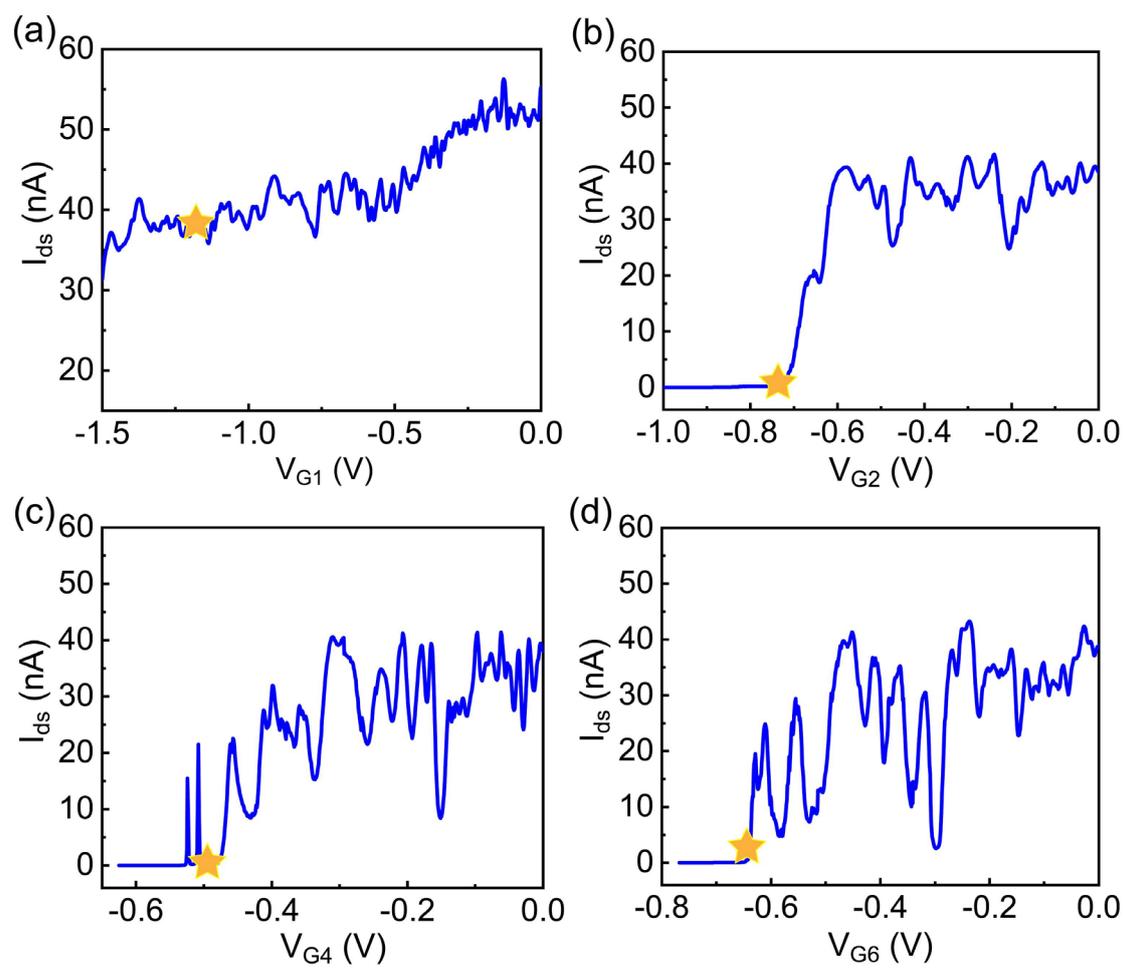

**Fig. 2** by Chen Y *et al.*

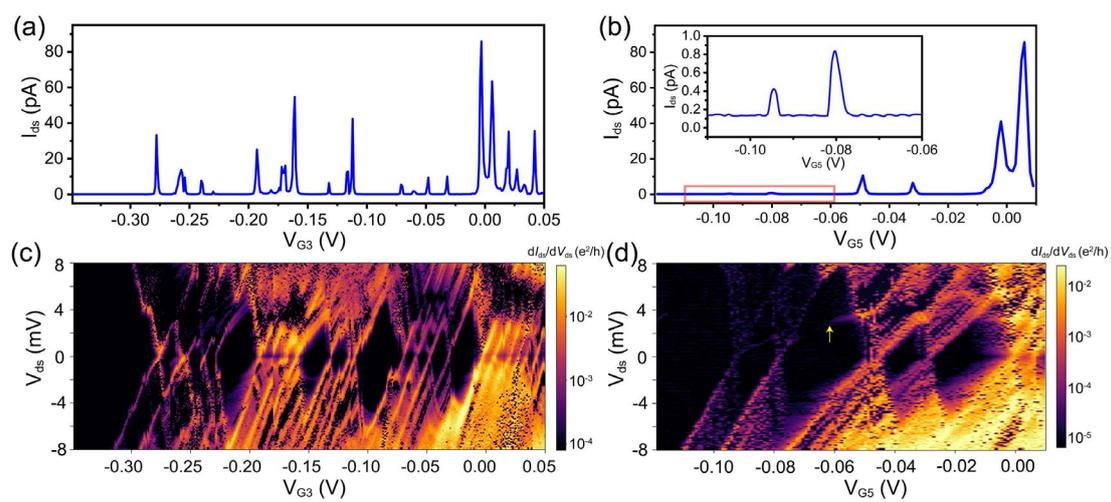

**Fig. 3 by Chen Y *et al*.**

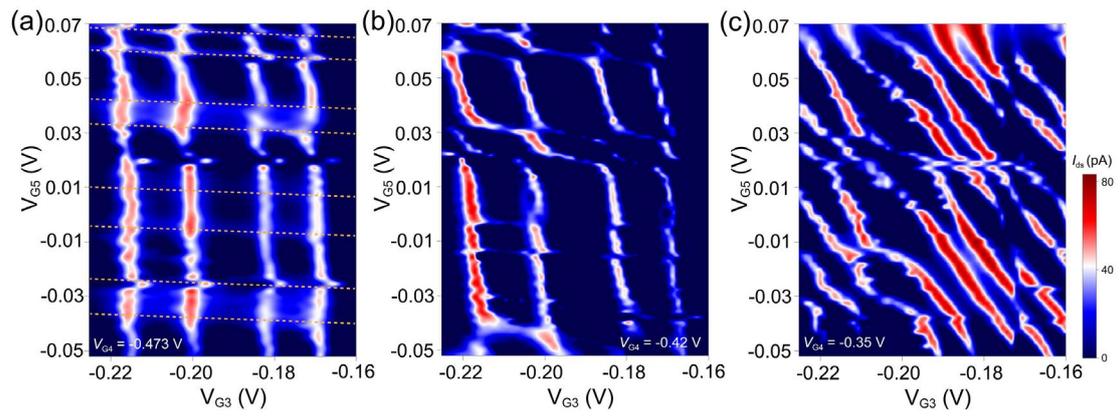

**Fig. 4 by Chen Y *et al*.**

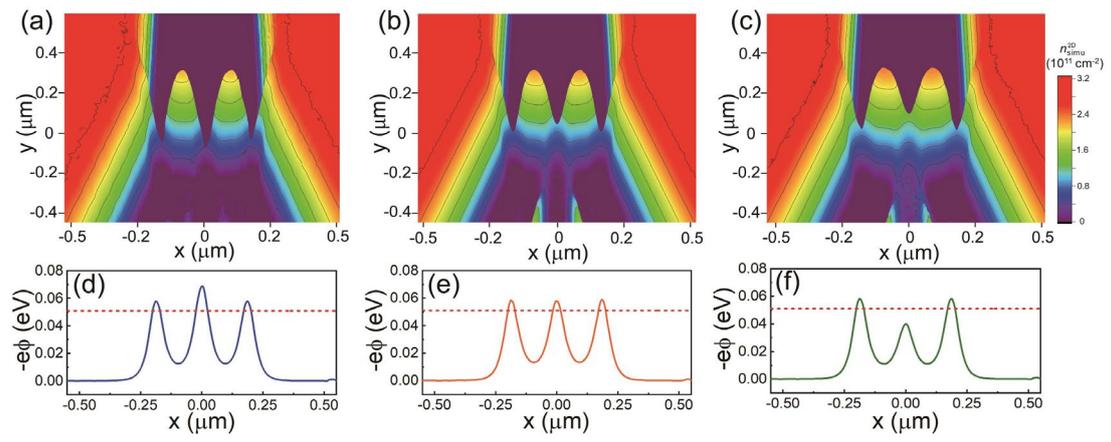

**Fig. 5** by Chen Y *et al*.